\documentclass[showpacs,amsfonts,notitlepage]{revtex4-1} 
\newcommand{\Dir}{\mathfrak{D}}
\newcommand{\AF}{\mathfrak{F}}
\newcommand{\ve}[1]{\bm{#1}} 
\newcommand{\eref}[1]{Eq.~(\ref{#1})} 
\newcommand{\Eref}[1]{Equation~(\ref{#1})} 
\newcommand{\tit}[1]{``#1,''}
\newcommand{\rld}{\iota}
\newcommand{\corr}{\gamma}
\newcommand{\dd}{\eta}
\newcommand{\av}[1]{\bar{#1}}
\newcommand{\GF}{\mathfrak{T}}
\newcommand{\qs}[1]{\tilde{#1}}
\newcommand{\Line}{\mathcal{L}}
\newcommand{\sref}[1]{Sec.~\ref{#1}}
\newcommand{\fref}[1]{Fig.~\ref{#1}}
\newcommand{\erefs}[2]{Eqs.~(\ref{#1},~\ref{#2})}

\usepackage{bm}
\usepackage{graphicx}
\begin{document}
%\sloppy
\title[Extension of Dirac's chord method]%
{Extension of Dirac's chord method to the case of a nonconvex set 
by use of quasi-probability distributions}%

\author{Alexander Yu.\ Vlasov}%

\email{qubeat@mail.ru, alexander.vlasov@pobox.spbu.ru}%

\affiliation{Federal Radiology Center (IRH)\\
197101, Mira Street 8, St.-Petersburg, Russia\\}
\affiliation{A. Friedmann Laboratory for Theoretical Physics\\
191023, Griboedov Canal 30/32, St.-Petersburg, Russia}%

\begin{abstract}
The Dirac's chord method may be suitable in different areas of physics for the
representation of certain six-dimensional integrals for a convex body using 
the probability density of the chord length distribution.
For a homogeneous model with a nonconvex body inside a medium with 
identical properties an analogue of the Dirac's chord method may be obtained,
if to use so-called generalized chord distribution. The function is defined 
as normalized second derivative of the autocorrelation function.
For nonconvex bodies this second derivative may have negative values 
and could not be directly related with a probability density. 
An interpretation of such a function using alternating sums of probability 
densities is considered. Such quasi-probability distributions may be used for 
Monte Carlo calculations of some integrals for a single body of arbitrary 
shape and for systems with two or more objects and such applications are  
also discussed in this work. 
\end{abstract}

\maketitle

%\ams{65C20, 65C05, 60D05, 60K40}

%pacs
%02.50.-r Probability theory, stochastic processes, and statistics
%02.50.Ng Distribution theory and Monte Carlo studies
%02.70.Tt Justifications or modifications of Monte Carlo methods  
%02.70.Rr General statistical methods (in computing) 
%02.30.Cj Measure and integration  

%msc
%65C20 Models, numerical methods
%65C05 Monte Carlo methods   
%60D05 Geometric probability and stochastic geometry 
%60K40 Other physical applications of random processes
%28C05 Integration theory via linear functionals 
%60A10 Probabilistic measure theory

\section{Introduction}
\label{Sec:intro}

Let us consider an integral
\begin{equation}
\AF_{B_1}^{B_2}(\varphi) = \int_{B_2}\int_{B_1}%
\frac{\varphi(|\ve{r}_1 - \ve{r}_2|)}{4 \pi |\ve{r}_1 - \ve{r}_2|^2}%
 d{\mathbf V}_1 d{\mathbf V}_2,
\label{PntKrn}
\end{equation}
where $B_1$, $B_2$ are three-dimensional bodies,
$\ve{r}_1 = (x_1,y_1,z_1) \in B_1 $, $\ve{r}_2 = (x_2,y_2,z_2)
\in B_2$ are pair of points, 
$d{\mathbf V}_1 = d x_1 d y_1 d z_1$ and 
$d{\mathbf V}_2 = d x_2 d y_2 d z_2$.

Similar integrals are used in different physical applications, e.g. 
in the calculations with a {\em point-kernel} method in the radiation shielding 
and dosimetry \cite{RS,M5}.

If $B_1 = B_2 = B$ --- is the single {\em convex} body,
the {\em Dirac's chord method} \cite{Dir1} may be applied for the calculation of 
the particular case of the double integral \eref{PntKrn} over pairs of points 
in the convex body $B$ using the probability density of the chord length 
distribution, $\mu(l)$
\begin{equation}
\Dir_{B}(\varphi) = \int_B\!\int_B%
\frac{\varphi(|\ve{r}_1 - \ve{r}_2|)}{4 \pi |\ve{r}_1 - \ve{r}_2|^2}%
 d{\mathbf V}_1 d{\mathbf V}_2 =
\frac{S_B}{4}\int_0^{\infty}\!\!\!\!\!\mu(l)%
\Bigl(\int_0^l\!\!\int_0^r\!\!\!\varphi(x)d x\,d r \Bigr) d l.
\label{IntDir}
\end{equation}
Here vectors $\ve{r}_1$, $\ve{r}_2 \in B$ represent pair of 
points of the body $B$ and $S_B$ is the surface area of $B$. The
infinite upper limit of integration is written for simplicity in the right-hand 
side of \eref{IntDir} and similar equations below due to obvious property
$\mu(l) = 0$ for $l > l_{\max}$, where $l_{\max}$ is the maximal 
possible length of a chord. 

Such a formula may be used in analytical and numerical 
methods of the calculation of the integrals such as $\Dir_{B}(\varphi)$. 
A demonstrative advantage is the reduction of a six-dimensional integral to 
an easier expression such as \eref{IntDir}. It is possible to obtain a direct
analytical formula for the chord length distribution (CLD) for some 
bodies and it was initially used by Dirac {\em et al}, Ref.~\onlinecite{Dir2}. 

Analytical expressions may be found only for few simple shapes and it
is reasonable to consider application of \eref{IntDir} for numerical 
calculations of integrals, e.g. for Monte Carlo methods.  
Indeed, both the Monte Carlo method \cite{Met} and the Dirac's chord method
from very beginning were used for the solution of analogue problems of the particle 
transport. The possibility to get rid of the singularity $1/R^2$ in the left-hand side 
of \eref{IntDir} is important for the application of Monte Carlo methods 
and it may be also actual for \eref{PntKrn} with two neighboring or overlapping 
regions $B_1$ and $B_2$.

However, even the generalization of \eref{IntDir} for a single {\em nonconvex} body 
is not obvious, because a straight line may intersect the body few times
and an appropriate choice of a definition of CLD is not quite clear in such a case. 
There are three widely used nonequivalent constructions of CLD for a nonconvex
body \cite{Gil00,MRG03,MRD03,BR01,Str01,Gil02,Han03,GMR05}. All intervals 
of the same line inside of a nonconvex body may be considered as separate 
chords to produce {\em the multi-chord distribution} (MCD). 
It is also possible to calculate the sum of lengths of all such intervals 
to define {\em the one-chord distribution} (OCD). 

The third definition introduces {\em a generalized chord distribution} as 
the second derivative of the autocorrelation function divided on some
normalizer (e.g., $S_B/4$) \cite{BR01,Str01,Han03,TorLu93}. 
It is justified, because for a convex body such a formal expression 
is equal to the probability density for CLD. 
In a more general case such a definition is also
useful, because just the generalized chord distribution should be used 
in \eref{IntDir} for a nonconvex body $B$ instead of CLD \cite{SCLD1} 
and it is discussed below.
However, for some nonconvex bodies the function may be negative 
for certain ranges of argument \cite{Gil02,Han03}.

Methods of construction of such functions as alternating (in sign of terms) sums 
of probability densities are utilized in the presented paper. 
Such an approach provides a direct analogue of \eref{IntDir} for calculation of 
integrals for {\em nonconvex bodies} \cite{SCLD1}. An extension of this 
technique may be appropriate for treatment of a more difficult 
case \eref{PntKrn} with two different bodies \cite{SCLD2}.

{\em Plan of the paper.}
In \sref{Sec:Ray} is revisited a {\em ray method} as a facilitated analogue
of the Dirac's chord method. It produces an understanding physical
model and introduces simplified versions of some tools 
applied further for chords. In \sref{Sec:Eqs} some equations are collected 
which are useful further for discussion about applications
of the chord method in \sref{Sec:Chord}. The integral \eref{PntKrn}
with two bodies and a multi-body case are discussed in \sref{Sec:Mult}. 
Methods of applications of considered techniques for the
statistical (Monte Carlo) sampling are discussed mainly in sections
\ref{Sec:MCR}, \ref{Sec:MCC} and \ref{Sec:MultRay}, \ref{Sec:MultChord}.

\section{Ray method}
\label{Sec:Ray}

\subsection{Ray length distribution}
\label{Sec:RLD}

There is an analog of \eref{IntDir} with the probability density of 
the ray length distribution (RLD), $\rld(l)$, i.e. instead of a full chord 
only a ray (segment) is considered. It is drawn from a point 
inside the body to the surface. The points have the uniform distribution 
and the directions of the rays are isotropic.
It may be written in such a case
\begin{equation}
\Dir_{B}(\varphi) =
V_B\int_0^{\infty}\!\!\!\!\rld(l)\Bigl(\int_0^l\!\!\varphi(x)d x \Bigr) d l,
\label{IntRay}
\end{equation}
where $V_B$ is the volume of $B$. 
This expression could be considered as an intermediate step in the derivation of
\eref{IntDir} in Ref.~\onlinecite{Dir1} and might be simpler for explanation adduced
below.

Let us introduce a simple isotropic homogeneous model with particles emitted 
inside a convex body $B$ and traveling along straight lines. If absorption
of energy on the distance $l$ from a source is defined by $\varphi(l)$, 
the left-hand side of \eref{IntDir} or \eref{IntRay} with six-dimensional 
integral describes a fraction of energy absorbed inside the body.

On the other hand, the same value may be calculated using the distribution of
particle tracks (rays) inside the body. The part of energy, 
absorbed on a ray with a length $l$ is 
\begin{equation}
I_\varphi(l)=\int_0^l\varphi(x)d x
\label{ifi}
\end{equation}
and a fraction of
rays with the length $l$ is described by RLD $\rld(l)$. 
It concludes an informal visual explanation of \eref{IntRay}, because 
the total amount of emitted particles is proportional to the volume of $B$.

\begin{figure}[hbt]
\begin{center}
\includegraphics[scale=0.333]{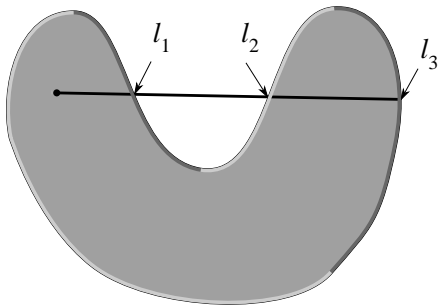}
\end{center}
\caption{Ray in nonconvex body}\label{Fig:ncray}
\end{figure}

The example with rays is also useful for the explanation of an appearance of
alternating sums of distributions. Let us consider the nonconvex body 
and the ray with three intersections with the boundary depicted in \fref{Fig:ncray}.

For each such ray instead of \eref{ifi} for the calculation of the energy
absorbed {\em inside} the nonconvex body an expression
\begin{equation}
\int_0^{l_1}\!\!\varphi(x)d x+\int_{l_2}^{l_3}\!\!\varphi(x)d x =
I_\varphi(l_1)-I_\varphi(l_2)+I_\varphi(l_3)
\label{ifi3}
\end{equation}
should be used,
where $I_\varphi$ is the antiderivative of $\varphi$ defined by \eref{ifi}.
It is possible to introduce few distributions $\rld_k(l)$ of distances 
from the source to $k$-th intersection and to write instead
of \eref{IntRay}
{\samepage
\begin{eqnarray}
\Dir_{B}(\varphi) &=&
V_B\sum_{k=1}^{k_{\max}} (-1)^{k+1}\int_0^{\infty}\!\!\!\!\rld_k(l)%
\Bigl(\int_0^l\!\!\varphi(x)d x \Bigr) d l \nonumber \\
&=& V_B\int_0^{\infty}\Bigl[\sum_{k=1}^{k_{\max}}(-1)^{k+1}\rld_k(l)\Bigr]%
\Bigl(\int_0^l\!\!\varphi(x)d x \Bigr) d l,
\label{IntRk}
\end{eqnarray}
}%
where $k_{\max}$ is the maximal number of intersections of a ray with the boundary
of $B$. The alternating sum in square brackets in \eref{IntRk} may be considered 
as a ``quasi-probability distribution'' $\qs{\rld}(l)$ and so \eref{IntRk}
may be rewritten to produce an analogue of \eref{IntRay}
\begin{equation}
\Dir_{B}(\varphi) =
V_B\int_0^{\infty}\!\!\!\!\qs{\rld}(l)\Bigl(\int_0^l\!\!\varphi(x)d x \Bigr) d l,
\quad \qs{\rld}(l) = \sum_{k=1}^{k_{\max}}(-1)^{k+1}\rld_k(l).
\label{SIntRay}
\end{equation}
More rigorous treatment may use so-called {\em signed measures (charges)} 
\cite{KolmFom} instead of term {\em quasi-probability distribution} used here.   
Some details may be found in Ref.~\onlinecite{SCLD1}.

The visual interpretation of equations for rays above is rather informal.
It was used understanding description with particles propagated along straight 
lines. Such a picture may create a wrong impression about impossibility to apply
considered methods to more difficult models with scattering. It is not so, because 
the only essential condition is the possibility to use in integrals like 
\eref{IntDir} expressions depending merely on $|\ve{r}_1 - \ve{r}_2|$.

An example of appropriate model is a convex body with absence of scattering,
but yet another case is an arbitrary body inside the medium with 
indistinguishable properties. The last case ensures possibility to apply 
\erefs{IntDir}{IntRay} and further generalizations  
to expressions with so-called {\em  build-up factors} used in dosimetry and 
radiation shielding to take into account the scattering \cite{RS,M5}. 
For the uniform and isotropic case such a build-up factor (for given energy) 
is again depending only on the distance from a point source.

\smallskip

Let us introduce polar coordinates in the second integral in left-hand side
of \eref{IntDir}. It makes the consideration more rigour \cite{Dir1}. 
Then for a convex body $B$
it is possible to write
\begin{equation}
\Dir_{B}(\varphi) =
\int_B d{\mathbf V}_1 
 \int_0^\pi\!\! \sin\theta d\theta\int_0^{2\pi}\!\!\! d\phi 
 \!\int_0^{l(\ve{r}_1,\theta,\phi)}\frac{\varphi(R)}{4\pi} d R, 
\label{IntVRPol}
\end{equation}
where with the preceding notation of \eref{IntDir} 
$R = |\ve{r}_1 - \ve{r}_2|$ together with $\theta, \phi$ are polar
coordinates of the vector 
$\ve{R} = \ve{r}_2 - \ve{r}_1$ and $l(\ve{r}_1,\theta,\phi)$
is the length of a ray from a point $\ve{r}_1$ with a direction given by 
the polar angles $\theta$ and $\phi$. 

%revision 1
A designation $d \Omega = \sin\theta\,d\theta\,d\phi$ for
the integration on the {\em solid angle}, $\Omega$ may be used for brevity 
\begin{equation}
\Dir_{B}(\varphi) =
 \int_B d{\mathbf V}_1 \int_{\mathbf S} \frac{d \Omega}{4\pi}
\!\int_0^{l(\ve{r}_1,\mathbf\Omega)}\!\!\!\!\varphi(R) d R, 
\label{IntVRS}
\end{equation}
where $l(\ve{r}_1,\mathbf\Omega)$ is the length of a ray from a point $\ve{r}_1$, i.e. 
$\mathbf\Omega$ denotes direction, represented earlier via $\theta$, $\phi$. 

\Eref{IntRay} may be now derived, if to take into account normalizing multipliers 
$V_B$ (volume of body $B$) and $4\pi$ (area of surface of unit sphere). 
It is explained below in \sref{Sec:Eqs}. Some additional technical discussion and 
references may be also found in Ref.~\onlinecite{SCLD1}.
Here is important to emphasize, that the ray in \eref{IntVRS} is not necessary a
particle trajectory, but a formal ``axis'' $\ve{R}$ of the integration on the 
variable $R$. 

Moreover, the formulas such as \eref{ifi3} or \eref{IntRk} with integrals 
on few disjoint intervals for a nonconvex body are also appropriate
here and so \eref{SIntRay} is valid. It justifies application of considered
methods for arbitrary isotropic uniform media, 
i.e. for models with scattering. 

The important example is a body (convex or nonconvex) inside an environment 
with identical or similar properties. In such a case the term in the left-hand 
side of \eref{IntDir} depends only on distance $|\ve{r}_1 - \ve{r}_2|$ even 
for points $\ve{r}_2$ near the boundary. For a convex body with straight tracks 
it is also true, but the environment does not matter, because trajectories of 
particles between two points inside the body may not fall outside the 
boundaries unlike the case with scattering.

\subsection{Method Monte Carlo with rays}
\label{Sec:MCR}

A useful application of \erefs{IntDir}{IntRay} is the Monte Carlo calculation
of integrals. There is an additional advantage for the calculation
of such integrals with many different $\varphi(l)$ for each body. In such a
case CLD or RLD for given body is calculated only once and used further
with different functions $\varphi(l)$. Functions, expressed via the definite 
integrals (single or double) of $\varphi(l)$ in right-hand side of the equations 
may be calculated either numerically or analytically.

The Monte Carlo sampling of a distribution is a standard procedure and 
may be visually represented as some {\em histogram}. The space between
zero and the maximal possible length is divided on $n$ {\em bins}, i.e.
sections $l_j \le l < l_j + \Delta l$, $j = 0,\ldots,n-1$ and during simulation for 
each step an amount of ``hits'' in an appropriate bin is increased by one. 
For the equal size $\Delta l$ of all sections the index $j$ of a bin is simply 
the integer part of $l/\Delta l$ and the tracing of such a data in the Monte Carlo 
simulations is fairly fast and useful procedure.

For the application of \eref{SIntRay} it is possible instead of
construction of $k_{\max}$ different distributions to 
create $\qs{\rld}(l)$ at once.
If a ray intersects boundary in few points it is necessary to consider
intervals from the origin to all points of intersection. For the length of each
interval with odd index (first, third, etc.) 
it is necessary to add unit to number of hits in a bin, 
but for interval with even index it is necessary to subtract unit from 
a number in the relevant bin. 

Such a method describes the Monte Carlo algorithm for the generation of 
the function $\qs{\rld}(l)$. More difficult algorithms for
quasi-probability distributions of chords are discussed below. 
However, it is reasonable at first to recollect some concepts for the
explanation, why such algorithms are relevant with alternative definitions 
via derivatives of the autocorrelation function.

\section{Helpful analytical equations}
\label{Sec:Eqs}

There are few functions related with presented models. It was 
already mentioned the chord length (distribution) density $\mu(l)$, the
ray length (distribution) density $\rld(l)$, and the autocorrelation function, 
denoted further as $\corr(l)$. It is also convenient to consider the probability
density of the distance distribution (DD) $\dd(l)$. There are important relations 
between these functions 
\cite{Gil00,MRG03,MRD03,BR01,Str01,Han03,TorLu93,SCLD1,Kel71,Maz03}
\begin{eqnarray}
  \mu(l) & = & \frac{\av{l}}{V_B}\corr''(l),
 \label{mu2cor} \\
 \mu(l) & = & -\av{l}\,\rld'(l)  
 \label{mu2rld} \\
 \rld(l) &=& -\frac{1}{V_B}\corr'(l),
 \label{rld2cor}\\
 \dd(l) &=& \frac{4\pi l^2}{V_B^2}\corr(l),
\label{dd2cor}
\end{eqnarray}
where $V_B$ is the volume of body $B$ and $\av{l}=\int_0^\infty{l\,\mu(l)d l}$
is the average chord length, that may be found for a convex body from a widely 
used Cauchy relationship \cite{Dir1,Kel71,Ken,San,Cau}
\begin{equation}
 \av{l} = 4\frac{V_B}{S_B}.
\label{Cau}
\end{equation}
The autocorrelation function $\corr(l)$ is defined here for body with density 
$\rho(\ve{r})=1$ for $\ve{r} \in B$ as 
\begin{equation}
 \corr(\ve{r}) = 
 \int_B \rho(\ve{r}_1) \rho(\ve{r}_1+\ve{r}) d \mathbf{V}_1, \quad
 \corr(l) = \frac{1}{4\pi l^2}\int_{|\ve{r}|=l} \corr(\ve{r}) d\Omega,
\label{corr}
\end{equation}
i.e. $d \mathbf{V}_1 = dx_1 dy_1 dz_1$, $\ve{r}_1 = (x_1,y_1,z_1)$ and
$\corr(l)$ is the average of $\corr(\ve{r})$ on a sphere with radius $l$.
Definition of $\corr$ here is lack of $1/V_B$ multiplier in comparison with 
some other works \cite{SCLD1} and it causes an insignificant difference in few 
equations.
In fact, only the property \eref{dd2cor} is used further and the formal 
definition \eref{corr} is presented here for completeness.

The probability density of the distance distribution $\dd(l)$ is easily defined for 
convex, nonconvex cases,
and also for a system of two bodies. For the explanation of relations between 
derivatives of $\corr$ in \erefs{mu2cor}{rld2cor} it is convenient 
to start with the expression 
\begin{equation}
\frac{1}{V_B^2}\Dir_{B}(\varphi) = \frac{1}{V_B^2}\int_B\int_B%
\frac{\varphi(|\ve{r}_1 - \ve{r}_2|)}{4 \pi |\ve{r}_1 - \ve{r}_2|^2}%
 d{\mathbf V}_1 d{\mathbf V}_2 =
\int_0^{\infty}\frac{\varphi(l)}{4 \pi l^2}\, \dd(l) d l.
\label{IntDD}
\end{equation}
It may be explained using a statistical approach 
convenient here due to discussion on the Monte Carlo sampling. 
The left-hand side of \eref{IntDD} may be considered as an average of 
a function $\Phi(R) = \varphi(R)/(4 \pi R^2)$ of a variable  
$R=|\ve{r}_1 - \ve{r}_2|$ defined on a space $B \times B$. 

The multiplier $V_B^2$ is a measure (6D volume) of
the six-dimensional space $B \times B$ and the division on this value is
due to averaging. On the other hand, the standard correspondence 
\cite{SkorPro} of a stochastic average with a {\em mathematical expectation} 
$\mathbf{E}$ makes possible to use the formula
\begin{equation}
\mathbf{E}\Phi(R) = \int \Phi(l)\, d_lF_R(l),
\label{MExp}
\end{equation}
where $F_R(l)$ is the cumulative distribution function of a random variable $R$ and 
$d_l F_R(l)$ denotes probability density of $R$, but in the considered case 
it is just the density of the distances distribution (DD) defined earlier 
$\dd(l) d l = d_l F_R(l)$. Really, $R$ is the distance between two points 
$\ve{r}_1$, $\ve{r}_2$ with independent uniform distributions and 
double integral in \eref{IntDD} corresponds to averaging on $B \times B$ 
for such a pair.

\Eref{IntDD} may be rewritten due to relation \eref{dd2cor} as 
\begin{equation}
\Dir_{B}(\varphi) = 
\int_0^{\infty}\!\!\!\varphi(l)\, \corr(l) d l.
\label{IntCor}
\end{equation}

After integration by parts it is possible to obtain from \eref{IntCor}
\begin{equation}
\Dir_{B}(\varphi) =
\int_0^{\infty}\!\!\!\![-\corr'(l)]\Bigl(\int_0^l\!\!\varphi(x)d x \Bigr) d l.
\label{IntCorI}
\end{equation}
For a convex body \eref{IntCorI} is in agreement with \erefs{IntRay}{rld2cor}.

\Eref{IntRay} may also be proven using an analogue of the statistical approach 
discussed above. A detailed proof may be found elsewhere \cite{SCLD1} and is only 
briefly sketched here. It is possible to consider \eref{IntVRS}
as an averaging on the five-dimensional space
of rays, represented as product $B \times {\mathbf S}$ of the body $B$ on 
the unit sphere ${\mathbf S}$. It is necessary to use for normalization
the volume $V_B$ of $B$ multiplied on $4 \pi$, the surface area of the unit sphere. 
In such a case \eref{IntRay} may be considered as an analogue of
\eref{MExp} for the mathematical expectation of some function depending
on the length of a ray.

\smallskip

It is possible to derive an equivalent of \eref{rld2cor} for {\em a nonconvex body} 
with $\qs{\rld}(l)$ introduced in \eref{SIntRay} if to use generalized 
functions and derivatives. The idea of a generalized function 
is convenient also for the further work with integrals such as $\Dir(\varphi)$.

The {\em generalized function (distribution)} is defined \cite{KolmFom} 
as {\em the continuous linear functional} $\GF(\phi)$ on the space of {\em test 
functions} $\phi$. Usual integrable function $\psi$ may be associated with 
a functional $\GF_\psi$ defined for the test function $\phi(x)$ as
\begin{equation}
  \GF_\psi(\phi) = \int_{-\infty}^\infty \psi(x) \phi(x)  d x. 
\label{Tpsi}
\end{equation}
On the other hand, $\Dir_B : \varphi \to \Dir_B(\varphi)$ is also a linear 
functional on a test function $\varphi$ and may be considered as some
generalized function on $(0,\infty)$ defined by given body $B$. 
A topology on the space of test functions and continuity\cite{KolmFom}, 
i.e. $\Dir_B(\varphi_k) \to \Dir_B(\varphi)$ for 
$\varphi_k \to \varphi$,  are not discussed here.

It is often used as a simplified notation $\psi$ instead of $\GF_\psi$ for
such a {\em regular} generalized function \eref{Tpsi} \cite{KolmFom}. 
In such a case \eref{IntCor} may be rewritten simply as \(\Dir_B = \corr_B\). 

The {\em generalized derivative} \cite{KolmFom} is defined as functional
\begin{equation}
\GF'(\phi) = -\GF(\phi').
\label{GenDer}
\end{equation}
Due to \eref{IntRay} it is possible for convex body $B$ to write 
$\Dir_B' = -V_B\,\rld_B$ and \eref{SIntRay} for an arbitrary body ensures
$\Dir_B' = -V_B\,\qs{\rld}_B$.

\section{Chord method}
\label{Sec:Chord}

\subsection{Chord length distribution}
\label{Sec:CLD}

For a convex body \eref{IntDir} may be rewritten with generalized functions
and derivatives as $\Dir_B'' = (S_B/4) \mu_B$. Here generalized functions
may be more appropriate, because due to the expression with the second derivative 
CLD is not a regular function even if DD is not smooth in some 
points.

\begin{figure}[hbt]
\begin{center}
\includegraphics[scale=0.333]{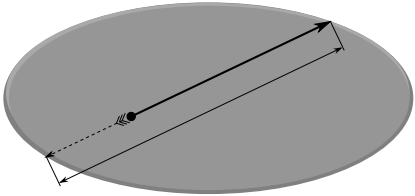}
\end{center}
\caption{Consideration of all possible rays along a chord}\label{Fig:ray2chord}
\end{figure}

Formally, for a convex body an expression with an additional integration along 
a chord appears due to a rearrangement of the integral \eref{IntVRPol} and 
consideration of all possible rays with origins along the same chord \cite{Dir1} 
(see~\fref{Fig:ray2chord}).

If to use a compact notation $\int d\Line$ for the formal integration
on a four-dimensional space of straight lines \cite{Ken,San}, it is possible
to rewrite \eref{IntVRS} after such a rearrangement as
\begin{equation}
\Dir_B(\varphi) = \int \frac{d{\Line}}{4 \pi} 
 \int_0^{l(\mathcal L)}\!\!\!\int_0^r\!\!\varphi(x)d x\,d r,
\label{IntL}
\end{equation}
where $l(\mathcal L)$ is the length of a chord produced by the intersection of
the straight line $\mathcal L$ with the convex body $B$. \Eref{IntL}
is an analogue of a familiar expression used for the derivation 
of the Dirac's chord method (see Eq.~(1.5) in Ref.~\onlinecite{Dir1}).

In such a case there is a double integral along a chord due to the 
additional integration on sources of rays
\begin{equation}
I^{(2)}_\varphi(l)=\int_0^l\!\!\int_0^r\!\!\varphi(x)d x\,d r.
\label{iifi}
\end{equation}
For a nonconvex body and few chords, i.e. $n$ intervals of intersections 
$[x_{2k},x_{2k+1}]$, $k = 0, \ldots n-1$ of the body by the same straight line,
it is necessary to include only the integration on both ``source'' points 
$r$ and ``target'' points $r_2 = r + x$ inside these intervals.
Using rather technical calculation (see Ref.~\onlinecite[Sec 3.3]{SCLD1}) it is possible 
to obtain instead of \eref{iifi} the more difficult expression 
{\samepage
\begin{eqnarray}
  I^{(2)}_{\varphi}(x_0,\ldots,x_{2n-1}) &=&
  \sum_{k=0}^{n-1}I^{(2)}_{\varphi}(x_{2k+1}-x_{2k}) \nonumber \\
  &+&     \sum_{k=1}^{n-1}\sum_{j=0}^{k-1}
     [I^{(2)}_{\varphi}(x_{2k+1}-x_{2j}) + I^{(2)}_{\varphi}(x_{2k}-x_{2j+1})]
   \nonumber \\
  &-& \sum_{k=1}^{n-1}\sum_{j=0}^{k-1}
     [I^{(2)}_{\varphi}(x_{2k+1}-x_{2j+1}) + I^{(2)}_{\varphi}(x_{2k}-x_{2j})].
\label{iifsum1}
\end{eqnarray}
It includes all $n (2n-1)$ possible ordered pairs $x_k - x_j$
with indexes  $0 \le j < k \le 2n-1$ and may be rewritten as
\begin{equation}
  I^{(2)}_{\varphi}(x_0,\ldots,x_{2n-1}) =
    \sum_{k=1}^{2n-1}\sum_{j=0}^{k-1}(-1)^{k-j+1}I^{(2)}_{\varphi}(x_k-x_j).
\label{iifsum2}
\end{equation}
}

\begin{figure}[hbt]
\begin{center}
\includegraphics[scale=0.333]{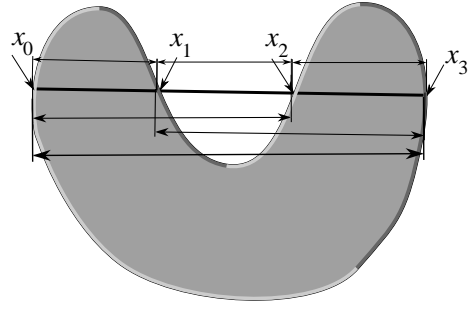}
\end{center}
\caption{Chords in nonconvex body and six possible segments}\label{Fig:ncxord}
\end{figure}

E.g., for two intersections there are six terms (see \fref{Fig:ncxord})
\begin{eqnarray*}
  &&I^{(2)}_{\varphi}(x_0,\ldots,x_3) 
 = I^{(2)}_{\varphi}(x_1-x_0) + I^{(2)}_{\varphi}(x_3-x_2) \\
  &&\qquad + I^{(2)}_{\varphi}(x_3-x_0) + I^{(2)}_{\varphi}(x_2-x_1) 
   -I^{(2)}_{\varphi}(x_2-x_0) - I^{(2)}_{\varphi}(x_3-x_1).
\end{eqnarray*}

Let us rewrite the integral \eref{IntL}
\begin{equation}
\Dir_B(\varphi) = \int \frac{d\Line}{4 \pi} 
 I^{(2)}_\varphi(\Line),
\label{IntLI}
\end{equation}
where for a convex body due to \eref{iifi} 
$I^{(2)}_\varphi(\Line) = I^{(2)}_\varphi(l_{\Line})$. 
The same expression also may be used for a nonconvex body if to denote
$I^{(2)}_\varphi(\Line) = I^{(2)}_\varphi(x^{\Line}_0,\ldots,x^{\Line}_{2n-1})$, 
where $x^{\Line}_0, \ldots, x_{2n-1}^{\Line}$
designate all intersections of the straight line $\Line$ with the boundary of $B$.

On the other hand, $I^{(2)}_\varphi(\Line)$ may be expressed
as a sum \eref{iifsum2} with all possible (ordered) pairs of points. 
It is possible to rewrite \eref{IntLI} for the nonconvex case
\begin{equation}
\Dir_B(\varphi) = \int \frac{d\Line}{4 \pi} 
    \sum_{k=1}^{2n-1}\sum_{j=0}^{k-1}
    (-1)^{k-j+1}I^{(2)}_{\varphi}(x^{\Line}_k-x^{\Line}_j).
\label{IntLIkj}
\end{equation}
The situation is similar with expressions for rays in a nonconvex
body \erefs{IntRk}{SIntRay}. 
Let us denote as $\mu_{jk}(l)$ probability densities of 
distributions of lengths $l_{jk} = x^{\Line}_k-x^{\Line}_j$ produced by $n(2n-1)$ 
ordered pairs $(x^{\Line}_j,x^{\Line}_k)$ on a line $\Line$. 

If to introduce 
\begin{equation}
  \qs{\mu}_{tot}(l) = 
   \sum_{k=1}^{2n-1}\sum_{j=0}^{k-1}(-1)^{k-j+1}\mu_{jk}(l),
 \quad \qs{\mu} = \qs{m}^{-1}\qs{\mu}_{tot}(l),
\label{SCLD}
\end{equation}
where $\qs{m}$ is the normalization 
\begin{equation}
 \qs{m} = \int_0^\infty\qs{\mu}_{tot}(l)d l,
\label{nSCLD}
\end{equation}
it is possible to write an analogue of \eref{IntDir} for a nonconvex body $B$
\begin{equation}
\Dir_{B}(\varphi) 
 = \qs{s}_B\int_0^{\infty}\!\!\!\!\qs{\mu}(l)I_{\varphi}^{(2)}(l)d l
 = \qs{s}_B\int_0^{\infty}\!\!\!\!\qs{\mu}(l)%
\Bigl(\int_0^l\!\!\int_0^r\!\!\varphi(x)d x\,d r \Bigr) d l,
\label{SIntDir}
\end{equation}
where $\qs{s}_B$ is some constant. 

For a convex body $\qs\mu(l)=\mu(l)$, $\qs{s}_B = S_B/4$ 
and \eref{SIntDir} may be explained
using an idea with the averaging and the mathematical expectation \eref{MExp} 
already discussed for DD and RLD.
Let us consider an average of the function $f(\Line) = I^{(2)}_\varphi(l_{\Line})$ 
on the four-dimensional set $\Line[B]$ of all straight lines intersecting the 
body $B$
\begin{equation}
 \frac{1}{w_B}\int_{\Line[B]}\!\!I^{(2)}_\varphi(l_{\Line}) d\Line=
\int_0^{\infty}\!\!\!\!I_{\varphi}^{(2)}(l) \mu(l)d l,
\label{IntLav}
\end{equation}
where $w_B$ is a measure (4D volume) of $\Line[B]$. 
For a convex body it may be expressed as 
$w_B = \pi S_B$ due to a Cauchy relationship \cite{SCLD1,Ken,San,Cau,Mat,Helg} and 
after comparison of \eref{IntLav} with \eref{IntL} we obtain necessary coefficient 
$w_B/(4\pi) = S_B/4$ used in \eref{IntDir}.

For a nonconvex body there are $n(2n-1)$ distributions $\mu_{jk}(l)$ instead
of one and \eref{SIntDir} is obtained via the alternating sum \eref{IntLIkj} of 
these distributions and so formula $\qs{s}_B = \qs{m}_B^{-1} w_B/(4\pi)$ is valid 
with $w_B$ is a measure for a set of all straight lines intersecting $B$ 
and $\qs{m}_B$ is a constant used in definition of $\qs{\mu}(l)$ \eref{SCLD}.
The problem here is an absence of simple methods of a calculation $w_B$ 
and $\qs{m}_B$ for nonconvex bodies and so it may be convenient to consider 
yet another approach for finding $\qs{s}_B$.

It is possible to use an analogue of the relation in \eref{mu2rld}. The integration
by parts of \eref{SIntRay} for a nonconvex body produces
\begin{equation}
\Dir_{B}(\varphi) 
 = V_B\int_0^{\infty}\!\!\!\![-\qs{\rld}'(l)]%
\Bigl(\int_0^l\!\!\int_0^r\!\!\varphi(x)d x\,d r \Bigr) d l
\label{SIntDirIP}
\end{equation}
and after the comparison with \eref{SIntDir} it is possible to write
\begin{eqnarray*}
- V_B\qs{\rld}'(l) &=& \qs{s}_B\qs{\mu}(l) \\
- V_B\int_0^{\infty}\!\!l\,\qs{\rld}'(l)d l 
 &=& \qs{s}_B\int_0^{\infty}\!\!l\,\qs{\mu}(l)d l \\
 V_B\int_0^{\infty}\!\!\qs{\rld}(l)d l &=& \qs{s}_B\av{l}
 \int_0^{\infty}\!\!\qs{\mu}(l)d l,
\end{eqnarray*}
where by definition 
$\av{l}_B = \int l\qs{\mu}_B(l) d l/\!\int\qs{\mu}_B(l) d l$
and due to normalization of $\qs{\rld}(l)$ and $\qs{\mu}(l)$ 
\begin{equation}
 \qs{s}_B = V_B/ \av{l}_B.
\label{qss}
\end{equation}
For a convex body $\av{l}_B$ is {\em the average chord length}. 
For a nonconvex body it is equal to the average chord length for the multi-chord 
distribution (MCD) mentioned earlier, because sums of lengths of all 
intervals in two last terms of \eref{iifsum1} compensate each other \cite{SCLD1}. 
It is clarified below in \sref{Sec:MCC} about Monte Carlo simulations.  

In fact, the Cauchy relationship \eref{Cau} for the average 
chord length for MCD is proved for a broad class of nonconvex
bodies \cite{MRG03} and so it is also possible to write due to 
\eref{qss} in such a case
\begin{equation}
 \qs{s}_B = S_B/4.
\label{qS4}
\end{equation}

For numerical methods using \eref{qss} with $\av{l}_B$ sometimes
may be preferable. 
It is instructive to consider an example with so-called voxel 
presentation of a body as a decomposition on small cubes or parallelepipeds. 
In such a case the surface is not smooth and a problem with correct approximation
of surface area may not be resolved even for a formal limiting case with 
cubes of arbitrary small dimensions, e.g. for a sphere such a limit 
is $6 \pi r^2$ instead of surface area $4 \pi r^2$.

\subsection{Method Monte Carlo with chords}
\label{Sec:MCC}

Let us consider some questions of the Monte Carlo generation of the quasi-probability 
distribution $\qs{\mu}(l)$. For each straight line with $n > 1$ intervals
inside a body $B$ it is necessary to consider $2n$ points of intersection with
the boundary of $B$. 
Tangent points should be counted twice. 
%but such degenerated cases are not discussed further for simplicity. 
Let us
mark all points by real numbers $x_k$, $k = 0,\ldots,2n-1$, there $x_0 = 0$ 
and other $x_k$ denote distances along a given straight line, i.e. 
$x_k = |\ve{r}_k-\ve{r}_0|$, where $\ve{r}_k$ denote positions
of $2n$ points of intersections in the three-dimensional space. 

It is clear from further consideration that it is possible to use opposite 
order of points $\ve{r}_{k} \leftrightarrow \ve{r}_{2n-k-1}$ and so 
``$\pm$~orientation'' of a line does not matter, i.e. two opposite directions
could not be distinguished. Anyway, in applied calculations it is often more 
convenient to use {\em directed lines}. Standard algorithms of the generation 
of uniform isotropic (pseudo)random sequences of straight lines should be 
discussed elsewhere.

Let us discuss the procedure of the construction of $\qs{\mu}(l)$ for the given line. 
If the line intersects a body $n$ times,
it is necessary to consider set of $2n$ numbers $x_k$ defined above
and to calculate lengths 
$l_{jk} = (x_k - x_j)$ for all $j, k: 0 \le j < k \le 2n-1$. 

For given $l_{jk}$ a number in the relevant bin should be {\em increased} 
by unit for {\em odd} $k-j$ and {\em decreased} by unit otherwise, 
i.e. if $k-j$ is {\em even}.
For $2n$ indexes there are $1 + 2 + \cdots + (2n-1) = n(2n-1)$ ordered pairs.
Between them $n$ pairs $(x_{2k},x_{2k+1})$, $k = 0,\ldots,n-1$
represent usual chords lying completely inside the body and they have 
positive contributions.

Remaining $n(2n-2)$ pairs are not forming 
intervals entirely belonging to the body and may be divided in two equal groups.  
There are $n(n-1)$ pairs with positive contribution, 
i.e. $(x_{2j},x_{2k+1})$ or $(x_{2j+1},x_{2k})$, 
$0 \le j < k \le n-1$.
For other $n(n-1)$ pairs, i.e. $(x_{2j},x_{2k})$ or $(x_{2j+1},x_{2k+1})$,
$0 \le j < k \le n-1$, 
numbers in appropriate bins should be decreased.

It is also clear from such a representation why sums of lengths
of the pairs in two last groups compensate each other:
\[
(x_{2k+1} - x_{2j}) + (x_{2k} - x_{2j+1}) = 
(x_{2k} - x_{2j}) + (x_{2k+1} - x_{2j+1}). 
\]
So sum of lengths alternating in signs is equal with summation of 
{\em only} $n$ positive contributions due to chords of the straight line 
inside the body.

There is also additional subtlety with normalization. 
For each straight line the total increase of values
in all affected bins is 
\[\Delta N_{tot} = n + n(n-1) - n(n-1) = n.\]
So there are two different counters: the number of
lines $N_l$ and the sum of numbers in all bins $N_{tot} \ge N_l$,
which is equivalent with a total number of separate chords lying 
completely inside the body.
The (quasi-probability) distribution should be divided in $N_{tot}$ for 
normalization. It is similar with the multi-chord distribution (MCD), 
because it is also normalized on the same number of separate chords $N_{tot}$.

It was shown above that total sum of all lengths while taking
into account signs is equal with sum of chords inside the body. 
But the normalization is the same as for MCD case and so
the formal averaging of the variable $l$ for the quasi-probability 
distribution $\qs{\mu}(l)$ constructed here is the same as the average
chord length for MCD that could be produced from the same set of straight lines. 
In a limit $N_l \to \infty$ it ensures equality of $\av{l}$
for both distributions already mentioned and used above in \eref{qS4}. 

The total number of lines $N_l$ corresponds to the normalization for OCD
case, when for each straight line only one ``aggregated'' chord, equivalent 
to union of all chords inside a nonconvex body, is considered. 
$N_l$ is also related with measure of set of all straight lines intersecting 
the considered body. Earlier in \eref{IntLav} this measure was denoted as $w_B$.
 
Yet another application of the both $N_{tot}$ and $N_l$ is the calculation of 
a constant $\qs{m}$ used earlier in \eref{SCLD}. It may be expressed as
a relation between $\qs{\mu}_{tot}$ (that is not normalized on
unit due to the contribution of lines intersecting the body more than one time) 
and $\qs{\mu}$.  So $\qs{m}$ is the limit of ratio between total number 
of chords $N_{tot}$ ({\em cf\/} MCD) and total number of straight lines 
$N_l$ ({\em cf\/} OCD)
\begin{equation}
 \qs{m} = \lim_{N_l \to \infty}N_{tot}/N_l.    
\label{Clim}
\end{equation}

\section{Multi-body case}
\label{Sec:Mult}

\subsection{Some equations with two different bodies} 
\label{Sec:MultInt}

This section is devoted to initial question about the calculation 
of the integral \eref{PntKrn} with two different bodies.
Here, it is also convenient to consider a simple model with particles
moving along straight lines in isotropic uniform medium and 
to use the interpretation of \eref{PntKrn} as a fraction of energy 
emitted in $B_1$ and absorbed in $B_2$.

\begin{figure}[hbt]
\begin{center}
\includegraphics[scale=0.333]{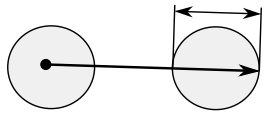}
\end{center}
\caption{Ray from $B_1$ with interval inside of $B_2$}\label{Fig:twobod}
\end{figure}

Let us consider a particle emitted in the first body with the straight trajectory
intersecting the second one (\fref{Fig:twobod}). If the law of absorption
is the same in both bodies and medium between them, it is possible
to describe amount of energy absorbed in the second body as 
\begin{equation}
\int_{a}^{b}\!\!\varphi(x)d x =  I_\varphi(b)-I_\varphi(a),
\label{ifi2}
\end{equation}
where $a$ and $b$ are distances from a source to two intersections
of second body by the ray and $I_\varphi$ is defined above in \eref{ifi}.

Calculation of \eref{PntKrn} for a convex $B_2$ would be
related with an analogue of \eref{IntVRPol}
\begin{equation}
\AF_{B_1}^{B_2}(\varphi) =
\int_{B_1} d{\mathbf V}_1 
 \int_{\theta_{\min}(\ve{r}_1,\phi)}^{\theta_{\max}(\ve{r}_1,\phi)}
 \!\!\sin\theta d\theta
 \int_{\phi_{\min}(\ve{r}_1)}^{\phi_{\max}(\ve{r}_1)}\!\!\! d\phi 
 \!\int_{a(\ve{r}_1,\theta,\phi)}^{b(\ve{r}_1,\theta,\phi)}
 \frac{\varphi(R)}{4\pi} d R, 
\label{IntVRPol2}
\end{equation}
where $\theta_{\min},\theta_{\max},\phi_{\min},\phi_{\max}$
describe angular limits of integrations for given point $\ve{r}_1$
and $a$, $b$ are radial distances for given point and direction.
It maybe simpler to use an analogue of \eref{IntVRS}
\begin{equation}
\AF_{B_1}^{B_2}(\varphi) =
\int_{B_1} d{\mathbf V}_1 
 \int_{{\mathbf S}(\ve{r}_1,B_2)}\! \frac{d \Omega}{4\pi}
\!\int_{a(\ve{r}_1,\mathbf\Omega)}^{b(\ve{r}_1,\mathbf\Omega)}\!\!\!\varphi(R) d R, 
\label{IntVRS2}
\end{equation}
where ${\mathbf S}(\ve{r}_1,B_2)$ is central projection from point $\ve{r}_1$
of body $B_2$ to surface of unit sphere. 
Yet, an alternative method of calculations is presented further and 
\erefs{IntVRPol2}{IntVRS2} are mentioned here rather for some clarification 
and comparison.

\subsection{Relation with methods for single body}
\label{Sec:MultOne}

For two nonconvex bodies expressions above could be even more difficult,
but it is possible to use a general principle to adapt already
developed approach with single body \cite{SCLD2}.
Let us choose both bodies as sources and
consider four integrals $\AF_{B_s}^{B_t}(\varphi)$, $s = 1,2$, $t=1,2$, i.e.
$\AF_{B_1}^{B_1} = \Dir_{B_1}$, $\AF_{B_2}^{B_2} = \Dir_{B_2}$, 
$\AF_{B_1}^{B_2} = \AF_{B_2}^{B_1}$. 
Each integral takes into account only particles emitted in
$B_s$ and absorbed in $B_t$.

The double integrals \erefs{PntKrn}{IntDir} comply with 
simple relations 
\begin{equation}
 \Dir_{B_1 \cup B_2}(\varphi) 
 = \AF_{B_1}^{B_1}(\varphi) + \AF_{B_1}^{B_2}(\varphi) + 
 \AF_{B_2}^{B_1}(\varphi)+ \AF_{B_2}^{B_2}(\varphi) 
\label{DirAFsum}
\end{equation}
and
\begin{equation}
 2 \AF_{B_1}^{B_2}(\varphi)= \Dir_{B_1 \cup B_2}(\varphi) -
 \Dir_{B_1}(\varphi) - \Dir_{B_2}(\varphi). 
\label{AF12Dir}
\end{equation}
So many equations with two bodies may be reduced to already
discussed case with the single body using union of 
these bodies $B = B_1 \cup B_2$. 
 
Here it is suggested that $B_1$ does not 
intersect $B_2$. For overlapping bodies the decomposition on
three parts: $B_1 \cup B_2$, $B_1 \setminus B_2$ and $B_2 \setminus B_1$
should be taken into account.
Instead of \eref{AF12Dir} in such a case a modified
expression may be used:
\begin{equation}
 2 \AF_{B_1}^{B_2}(\varphi)= \Dir_{B_1 \cup B_2}(\varphi) +
 \Dir_{B_1 \cap B_2}(\varphi) -
 \Dir_{B_1}(\varphi) - \Dir_{B_2}(\varphi). 
\label{AF3Dir}
\end{equation}

Due to such equations, the computational methods discussed above make possible
to find \eref{PntKrn} 
after separate calculations of three or four terms in \eref{AF12Dir} or \eref{AF3Dir}.
However, more direct approach discussed further is also useful and may be 
simply generalized for the case with many bodies.

A simpler case of two {\em disjoint bodies} is suitable for
almost straightforward modifications of Monte Carlo algorithms 
discussed above \cite{SCLD2}. 
Here \eref{DirAFsum} demonstrates that distributions obtained in
simulation may be divided in four parts (for each pair source-target)
without significant modification of algorithms for general
nonconvex body discussed above and it may be even more convenient 
for explanation than \eref{AF12Dir}. 

For two discontiguous bodies a union $B = B_1 \cup B_2$
formally is always nonconvex, because a line between a point in $B_1$ and
a point in $B_2$ lies partially outside the union.
So, here, consideration of quasi-probability distributions is especially justified.

For non-overlapping bodies with adjoining boundaries $B$ may be convex even 
with nonconvex $B_1$ or $B_2$. 
It corresponds to the case briefly mentioned below in a note about zones at the end
of \sref{Sec:MultChord}. It is enough to consider some body $B$ and formally
split it into two zones (parts) $B_1$ and $B_2$. Even for convex body such parts 
may be either convex or nonconvex. 
These subtleties do not affect on methods discussed here. Insignificant 
change is necessary only for a case with overlapping bodies due to contribution
of intersections with {\em nonzero volume} outlined in \eref{AF3Dir}.

\subsection{Application to calculations with rays}
\label{Sec:MultRay}

Source points uniformly distributed in both bodies with equivalent density 
are used for sampling with rays. 
All intersections of rays with boundaries of both bodies 
are checked and appropriate bins
are changed in four histograms $H_{st}(l)$ marked by indexes source-target. 
%The same histogram may be used for pair $H_{12}$ and $H_{21}$.

Joint consideration of all distributions lets to tackle a problem
with normalization. For the function $\qs{\rld}$ term 
``quasi-probability distribution'' could be justified due to normalization 
on unit integral and some relations with the probability density for length of 
rays in a convex body, but if to write an analogue of \eref{IntRay} for the
integral \eref{PntKrn}
\begin{equation}
\AF_{B_1}^{B_2}(\varphi) = 
W_{12}\int_0^{\infty}\!\!\!\!\qs{\rld}_{(12)}(l)
\Bigl(\int_0^l\!\!\varphi(x)d x \Bigr) d l,
\label{IntRay12}
\end{equation}
where $W_{12}$ is an unknown constant, it is simple to show that
$\qs{\rld}_{(12)}(l)$ may not be normalized for disjoint bodies,
because due to \eref{AF12Dir}
\[
 W_{12}\int_0^\infty\!\!\qs{\rld}_{(12)}(l)d l=
 (V_1+V_2)\int\!\!\qs{\rld}_{\cup}(l)d l -
 V_1\int\!\!\qs{\rld}_1(l)d l -
 V_2\int\!\!\qs{\rld}_2(l)d l = 0, 
\]
where integrals of all functions $\qs{\rld}_1(l)$, $\qs{\rld}_2(l)$ and 
$\qs{\rld}_{\cup}(l) = \qs{\rld}_{B_1\cup B_2}(l)$ are normalized on unit.

However, if to include all four densities as components in the single process
described by a (quasi-probability) distribution, introduced earlier
\begin{equation}
\qs{\rld}_{\cup}(l) = \sum_{s,t} \qs{\rld}_{(st)}(l),
\label{mrld}
\end{equation}
it is possible to consider $\qs{\rld}_{(st)}(l)$ as elements of
some matrix $\mathbf{\qs{\rld}}(l)$ with the common normalization. 
The same approach may be used for more than two bodies.

\subsection{Application to calculations with chords}
\label{Sec:MultChord}

The expression of \eref{PntKrn} via chord distributions also may
use similar principles \cite{SCLD2}. Here, it is also appropriate
to use the decomposition \eref{DirAFsum}. A straight line again determines
$2n$ values $x_0,\ldots,x_{2n-1}$, but boundaries of both
bodies must be marked appropriately. Each intersection should be refined as 
$x_k^s$ with additional index $s=1,2$ for $B_1$, $B_2$. 

Each pair $(x_j^s,x_k^t)$
already has two additional indexes $s$ and $t$ representing four possible 
combinations for two bodies and it produces distributions $\mu^{(st)}_{jk}(l)$ combined
with appropriate signs $(-1)^{k+j-1}$ into $\qs{\mu}_{(st)}(l)$, $s,t = 1,2$. 
It should be only mentioned that due to a symmetry for the chords  
``source'' and ``target'' bodies could be hardly distinguished. 
Due to such property it is reasonable to use only three separate 
histograms $H_{11}$, $H_{22}$ and $H_{12} + H_{21}$ 
and to define $\qs{\mu}_{\{st\}}(l)$ as a symmetric matrix. 

It may be directly generalized
for a case with $m$ bodies with $s,t = 1,\ldots,m$.
Advantages of application of discussed methods may
be illustrated by consideration of a domain with many different
bodies intersected by the variety of straight lines. It is possible
to calculate all $m^2$ integrals $\AF_{B_s}^{B_t}$ during
the same Monte Carlo simulation. 

Here only $m(m+1)/2$ integrals are different due to symmetry, 
but anyway it may be a big number. For medical applications
with $15-20$ objects (organs) it is the calculation of hundreds values 
in a single Monte Carlo run. In fact, speed up may be
even more critical due to possibility to split each body
into few zones. The subdivision may be necessary for taking into account 
a variation of the intensity of emitters in the different parts
of some objects.

There is a subtlety for the calculation with few zones: 
it is necessary formally to split each boundary between two zones into 
two coinciding surfaces. In such a case all equations above are valid,
but there are some intervals with zero length. Such intervals
may be simply omitted, because integration along them produces
zero values.   

\subsection{Analytical expressions for two bodies}
\label{Sec:MultEqs}

There are useful analogues of expressions discussed in the \sref{Sec:Eqs}
for the case with two bodies. A function $\dd_{(12)}(l)$ may be defined 
as the probability density of distances between a pair of points
uniformly distributed in first and second body, respectively. 
Analytical expressions written below may be used for the testing of
the Monte Carlo simulation and some clarification. 
Technical details may be found in Ref.~\onlinecite{SCLD2}.

The correlation function $\corr_{(12)}(l)$ is defined for two bodies
with unit densities $\rho_k(\ve{r})=1$ for $\ve{r} \in B_k$, $k = 1,2$ as
\begin{equation}
 \corr_{(12)}(\ve{r}) 
 = \int_{B_1} \rho_1(\ve{r}_1) \rho_2(\ve{r}_1+\ve{r}) d \mathbf{V}_1, \quad
 \corr_{(12)}(l) = \frac{1}{4\pi l^2}\int_{|\ve{r}|=l} \corr_{(12)}(\ve{r}) d\Omega.
\label{corr12}
\end{equation}

It is possible to derive the direct analogue of \eref{dd2cor} 
\begin{equation}
 {\dd}_{(12)} = \frac{4 \pi l^2}{V_1 V_2} \corr_{(12)}(l)
\label{dd2cor12}
\end{equation}
and to write the generalization of \eref{IntCor}
\begin{equation}
\AF_{B_1}^{B_2}(\varphi) =
\int_0^{\infty}\!\!\!\varphi(l)\, \corr_{(12)}(l) d l.
\label{IntCor12}
\end{equation}
 
Using integrations of \eref{IntCor12} by parts it is possible 
to express $\qs{\rld}_{(12)}$ and $\qs{\mu}_{(12)}$ as first
and second derivatives of correlation function $\corr_{(12)}(l)$,
respectively. It is similar with \eref{rld2cor} and \eref{mu2cor}.
If to use the normalization on the union of bodies suggested above and the
Cauchy relationship for the average chord length \eref{Cau}, it may be written 
\begin{eqnarray}
  \qs{\mu}_{(12)}(l) & = & \frac{4}{S_{B_1}+S_{B_2}}\corr''_{(12)}(l),
 \label{mu2cor12} \\
 \qs{\rld}_{(12)}(l) &=& -\frac{1}{V_{B_1}+V_{B_2}}\corr'_{(12)}(l).
 \label{rld2cor12}
\end{eqnarray}
Due to such normalization for the ray method an ``unknown constant'' in
\eref{IntRay12} may be chosen as
\begin{equation}
 W_{12} = V_{B_1}+V_{B_2}
\label{W12}
\end{equation}
and desired generalization of the Dirac's chord method for
the initial equation \eref{PntKrn} may be written finally as
\begin{equation}
\AF_{B_1}^{B_2}(\varphi) = 
\frac{S_{B_1}+S_{B_2}}{4}\int_0^{\infty}\!\!\!\!\qs{\mu}_{(12)}(l)%
\Bigl(\int_0^l\!\!\int_0^r\!\!\varphi(x)d x\,d r \Bigr) d l.
\label{IntDir12}
\end{equation}

\section{Conclusion}

A novel approach with the application of quasi-probability distributions 
(signed measures) to calculations of integrals such as \erefs{PntKrn}{IntDir} 
is advocated in presented work. It may be useful in many areas 
of physics. This paper is written with a purpose to present a fairly brief, 
but closed description of considered methods.
Additional technical details, proofs of some equations
together with appropriate links with theory of geometrical probabilities 
may be found elsewhere \cite{SCLD1,SCLD2}. 

It is shown, how models with ray and chord length distributions
suitable for a single convex body 
should be altered for a nonconvex case and multi-body systems. 
An essential new property of such extensions is the necessity to use instead 
of probability densities some functions, which sometimes do not satisfy 
the non-negativity condition.

Maybe such a counterintuitive ``negative probability'' produced certain difficulties 
and a delay in development and applications of these methods despite of high 
effectiveness of numerical algorithms based on ray and chord distributions.
On the other hand, quasi-probability distributions are rather common
in quantum physics after so-called Wigner function representation \cite{WigFun}
and Feynman wrote an essay about the concept of negative probability
with reasonable examples both in quantum and classical physics \cite{FeyNeg}.

In fact, the functions $\qs{\rld}(l)$ and $\qs{\mu}(l)$ do not necessarily
{\em directly} related with probability distributions and so should not 
cause some conceptual challenges. Appearance of 
negative values may be simply illustrated using \eref{ifi3} and \fref{Fig:ncray}. 
Here the ray $(0,l_3)$ includes a ray $(0,l_1)$ already taken into account and 
the interval $(l_1,l_2)$ outside of the body, that should not be counted at all. 

For the work with an interval $(l_2,l_3)$  expressions such as \eref{ifi3} 
were used, but it may be described in the standard probability theory.
If probability measures are known for sets 
$R_1 = A$, $R_2 = A \cup B$, $R_3 = A \cup B \cup C$, it is possible 
to write for $C$: ${\mathbf P}(C) = {\mathbf P}(R_3 \setminus R_2)
 = {\mathbf P}(R_3)-{\mathbf P}(R_2)$ and for $A \cup C$: 
${\mathbf P}(A \cup C) = {\mathbf P}(R_1) - {\mathbf P}(R_2)+{\mathbf P}(R_3)$. 

Overlapping sets, i.e. rays with the same origin, are used in construction of 
$\qs{\rld}(l)$ (see \fref{Fig:ncray}). Positive and negative
terms such as ${\mathbf P}(R_3)$ and $-{\mathbf P}(R_2)$, used for the calculation 
of the same ${\mathbf P}(C)$, affect two ranges of argument $\qs{\rld}(l)$. 
So for $l=R_3$ there is some 
positive gain, but for $l = R_2$ there is corresponding decrease and it may
produce negative values of $\qs{\rld}(l)$ for some intervals of $l$. 
An extra hit is added to some bin. A removal from another bin --- is an effort 
to compensate that, but it may produce a negative result.
 
The construction of $\qs{\rld}(l)$ is simpler, than the generalization of
the chord length distribution $\qs{\mu}(l)$, but a reason of appearance
of negative values in both cases is similar. An amount of terms in 
expressions for a ray grows linearly with respect to a number of intersections 
and for chord it is quadratic dependence. The structure of sets is also
more complicated for chords, but here alternating signs in formulas 
such as \eref{iifsum1} again correspond to an expression with unions and differences 
of some overlapped sets.

\end{document}